\documentclass{aa}
\usepackage{graphicx}
\graphicspath{{figures/}}
\usepackage{txfonts}
\usepackage{siunitx}
\newcommand{\dd}{\mathrm{d}}
\usepackage{multirow}
\usepackage[colorlinks=true, linkcolor=blue, citecolor=blue, filecolor=blue, urlcolor=blue]{hyperref}
\usepackage[nameinlink]{cleveref}
\crefname{section}{Sect.}{Sects.}
\crefname{equation}{Eq.}{Eqs.}
\crefname{figure}{Fig.}{Figs.}
\crefname{table}{Table}{Tables}
\crefname{appendix}{Appendix}{Appendices}
\Crefname{section}{Section}{Sections}
\Crefname{equation}{Equation}{Equations}
\Crefname{figure}{Figure}{Figures}
\Crefname{table}{Table}{Tables}
\Crefname{appendix}{Appendix}{Appendices}
\begin{document}

    \title{Three-dimensional continuum radiative transfer of polarized radiation in exoplanetary atmospheres}
    
    \titlerunning{3D radiative transfer in exoplanetary atmospheres}
    
    % \subtitle{subtitle}
    
    \author{M. Lietzow\and S. Wolf\and R. Brunngräber}
    
    \authorrunning{Lietzow et al.}
    
    \institute{Institute of Theoretical Physics and Astrophysics,
              Kiel University, Leibnizstr. 15, 24118 Kiel, Germany\\
              \email{mlietzow@astrophysik.uni-kiel.de}
              }
    
    \date{Received / accepted }

    % \abstract{}{}{}{}{}
    % 5 {} token are mandatory

    \abstract
    {Polarimetry is about to become a powerful tool for determining the atmospheric properties of exoplanets.
    For example, recent observations of the WASP-18 system allowed the polarized flux resulting from scattering in the atmosphere of WASP-18b to be constrained.
    To provide the basis for the interpretation of such observational results and for predictive studies to guide future observations, sophisticated analysis tools are required.}
    {Our goal is to develop a radiative transfer tool that contains all the relevant continuum polarization mechanisms for the comprehensive analysis of the polarized flux resulting from the scattering in the atmosphere of, on the surface of, and in the local planetary environment (e.g., planetary rings, exomoons) of extra-solar planets.
    Furthermore, our goal is to avoid common simplifications such as locally plane-parallel planetary atmospheres, the missing cross-talk between latitudinal and longitudinal regions, or the assumption of either a point-like star or plane-parallel illumination.}
    {As a platform for the newly developed numerical algorithms, we use the 3D Monte Carlo radiative transfer code POLARIS.
    The code is extended and optimized for the radiative transfer in exoplanetary atmospheres.
    We investigate the reflected flux and its degree of polarization for different phase angles for a homogeneous cloud-free atmosphere and an inhomogeneous cloudy atmosphere.
    Our results are compared with already existing results to verify the implementations.
    To take advantage of the 3D radiative transfer and to demonstrate the potential of the code, the impact of an additional circumplanetary ring on the reflected polarized flux is studied.
    Therefore, a simple ring model with water-ice particles is used and various inclination angles, optical depths and viewing angles are investigated.}
    {The considered test cases show a good agreement with already existing results.
    The presence of a circumplanetary ring consisting of small water-ice particles has a noticeable impact on the reflected polarized radiation.
    In particular, the reflected flux strongly increases at larger phase angles if the planetary orbit is seen edge-on because the considered particles tend to scatter forwards.
    In contrast, the degree of polarization decreases at these phase angles.}
    {We present a polarization radiative transfer tool in which all relevant contributions to the reflected polarized continuum flux are considered.
    In a case study, we investigated the impact of a planetary ring on the net polarization signal.}


    \maketitle
%
%-----------------------------------------------------------------------

\section{Introduction}
\label{sec:introduction}

    Since the discovery of the first extra-solar planet by \citet{Mayor_1995}, the number of detected exoplanets increased enormously to over \num{4000} (The Extrasolar Planets Encyclopaedia\footnote{\url{http://exoplanet.eu}}; \citealt{Schneider_2011}; from November 2020).
    The current generation of imaging instruments, for example the Gemini Planet Imager \citep[GPI;][]{Macintosh_2008} or the SPectro-Polarimetric High-Contrast Exoplanet Research \citep[SPHERE;][]{Buezit_2019}, potentially provide the opportunity to detect the infrared polarized signal from sufficiently bright (e.g., self-luminous) exoplanets.
    Furthermore, with modern polarimeters, such as the HIgh-Precision Polarimetric Instrument \citep[HIPPI;][]{Bailey_2015} or the POlarimeter at Lick for Inclination Studies of Hot jupiters 2 \citep[POLISH2;][]{Wiktorowicz_2015}, it is now possible to measure the polarized flux at a parts-per-million level.
    In addition, new high-accuracy polarimeters are going into operation soon, such as the High-Precision Polarimetric Instrument-2 \citep[HIPPI-2;][]{Bailey_2020}, or are in preparation, such as POLLUX \citep{Bouret_2018}.
    
    Recently, \citet{Bott_2018} reported observations of linear polarized radiation of the WASP-18 system, which harbors a massive planet (approximately \SI{10}{M_J}) orbiting close to its star with an orbital period of $p < \SI{1}{day}$.
    Although the measured polarization is dominated by the interstellar medium, the authors were able to set an upper limit of \SI{40}{ppm} ($99 \%$ confidence level) on the amplitude of a reflected polarized radiation planetary signal.
    Thus, they could rule out certain atmospheric models, such as optically thick atmospheres dominated by Rayleigh scattering clouds.

    Since the emitted radiation of solar-type stars can be assumed to be unpolarized \citep{Kemp_1987}, the reflected radiation of a planet is solely polarized due to scattering processes within its atmosphere.
    Different types and properties of atmospheric particles cause different characteristics in the polarization.
    Analyzing the reflected polarized radiation can therefore be a useful tool not only for detecting extra-solar planets, but also for characterizing their atmospheres.
    \citet{Hansen_Hovenier_1974} showed the benefits of this method by determining the cloud properties of the atmosphere of our neighboring planet Venus.
    With the previously mentioned instruments, detailed studies and characterization of exoplanetary atmospheres have come into reach.
   
    Meanwhile, various theoretical studies of linear polarization resulting from Rayleigh scattering \citep{Buenzli_2009} and scattering by water clouds \citep{Karalidi_2012}, circular polarization of cloudy exoplanets \citep{Rossi_2018a}, the determination of the cloud coverage of exoplanets \citep{Rossi_2018b}, and the linear polarization of scattered radiation from self-luminous exoplanets \citep{Stolker_2017} have been performed.
    These studies have in common that they either focus on specific polarization mechanisms or assume certain simplifications.
    Examples of common simplifications are the assumption of a locally plane-parallel planetary atmosphere, the often missing cross-talk between latitudinal and longitudinal regions (i.e., locally horizontal radiation transport) in the scattering process (not in the case of Monte Carlo simulations), black planetary surfaces (i.e., absorbing surfaces), the often neglected state of circular polarization, and the assumption of either a point-like star or plane-parallel illumination on the planetary atmosphere.
    
    Simulating the radiative transfer in a planetary atmosphere is a very complex problem since various parameters have an impact on the observed radiation.
    Furthermore, the planetary surface and its influence become relevant if the atmosphere is optically thin.
    The reflected (polarized) radiation caused by the surface depends on the material, such as land masses with or without biomass, or oceans.
    However, the reflected flux of an optically thick atmosphere is dominated by the wavelength-dependent scattering and absorption properties of the atmospheric particles.
    
    Our goal is to develop a polarization radiative transfer tool that considers all the relevant contributions to the continuum polarization signal due to various polarization mechanisms while avoiding simplification inherent to previous approaches.
    Thus, we provide a tool for the comprehensive analysis of the polarized flux reflected by extra-solar planets.
    This development is based on the publicly available 3D Monte Carlo radiative transfer code POLARIS\footnote{\url{http://www1.astrophysik.uni-kiel.de/~polaris}} \citep{Reissl_2016}.
    The code is already optimized to handle the full spectrum of state-of-the-art polarization mechanisms (e.g., scattering and/or thermal re-emission, line and/or continuum polarization).
    It is well tested and has been applied to a broad range of astrophysical models: molecular clouds \citep{Reissl_2017, Pellegrini_2019, Seifried_2020}, Bok globules \citep{Brauer_2016}, and protoplanetary disks \citep{Brauer_2019, Heese_2020, Brunngraeber_2020}.
    
    In \cref{sec:radiative_transfer_model} we define our radiative transfer model. The Stokes parameters that describe the intensity and state of polarization of the radiation are introduced in \cref{subsec:stokes_formalism}.
    In \cref{subsec:illuminating_source} the emission of photon packages by a spatially extended radiation source is briefly described, including our approach to reduce the run-time of the code by restricting the emission of photon packages.
    In \cref{subsec:planetary_model} the planetary model, its atmosphere, and its surface are briefly outlined.
    In \cref{sec:selected_test_cases} we verify the new routines and present the results for various atmospheric models including diffuse surface reflection.
    The results are compared to analytical solutions in the case of a simple diffuse reflecting sphere and already existing computations for various structured atmospheres.
    As an example case study (\cref{sec:circumplanetary_ring}), we investigate the presence of an additional circumplanetary ring and its impact on the reflected (polarized) flux.
    Finally, our results are summarized in \cref{sec:summary}.
    
%-----------------------------------------------------------------------

\section{Radiative transfer model}
\label{sec:radiative_transfer_model}

    We briefly introduce the Stokes parameters and the description of scattering that are used to model the radiation field and the radiative transfer.
    In order to adapt the existing radiative transfer code POLARIS to the specific requirements for handling radiation scattering in planetary atmospheres, both the illuminating source and the atmosphere have to be considered in detail afterwards.
    
%-----------------------------------------------------------------------
    
\subsection{Stokes formalism}
\label{subsec:stokes_formalism}
    
    The radiation field is represented by photon packages that are defined by their wavelength-dependent Stokes parameters.
    The Stokes parameters are used to determine the intensity, state, and degree of polarization \citep[see, e.g.,][]{Bohren_1983}; they are combined in a 4D vector $\vec{S} = (I, Q, U, V)^\mathrm{T}$, where $I$ is the total intensity, $Q$ and $U$ are the linear polarization, and $V$ the circular polarization.
    Furthermore, the quantities
    \begin{equation}
        P_\mathrm{l} = \frac{\sqrt{Q^2 + U^2}}{I}, \quad
        \tan( 2\gamma ) = \frac{U}{Q}, \quad
        P_\mathrm{c} = \frac{V}{I},
    \end{equation}
    describe the degree of linear polarization $P_\mathrm{l}$, the corresponding angle of linear polarization $\gamma$, and the degree of circular polarization $P_\mathrm{c}$, respectively.
    If the photon package scatters, the change in polarization is obtained by multiplying the incoming Stokes vector $\vec{S}_\mathrm{in}$ with a scattering matrix $\vec{F}(\Theta, \Phi)$, where $\Theta \in [0, \pi]$ and $\Phi \in [0, 2\pi]$ are the scattering angles.
    The resulting Stokes vector after scattering is then given by
    \begin{equation}
    \label{eq:stokes_scattering}
        \vec{S}_\mathrm{out} \propto \vec{F}(\Theta, \Phi) \cdot \vec{L}(\Phi) \cdot \vec{S}_\mathrm{in}.
    \end{equation}
    Here the rotation matrix $\vec{L}(\Phi)$ rotates the Stokes vector into different frames (e.g., into the scattering plane or into the observers frame) if the photon package is detected.
    We assume that the atmospheric particles are spherical and distinguish between Rayleigh scattering (particle size $r$ smaller compared to the wavelength $\lambda$) and the more general case of Mie scattering (arbitrary particle size).
    The general scattering matrix (or Müller matrix) for spherical particles has the following simplified structure \citep[e.g.,][]{Bohren_1983}:
    \begin{equation}
        \vec{F}(\Theta) =
        \begin{pmatrix}
        F_{11} & F_{12} & 0 & 0\\
        F_{12} & F_{22} & 0 & 0\\
        0 & 0 & F_{33} & F_{34}\\
        0 & 0 & -F_{34} & F_{44}
        \end{pmatrix}.
    \end{equation}
    
    In addition to their wavelength-dependent Stokes vector, the photon packages are characterized by their point of emission or last point of interaction as well as the direction of their propagation through the model space.
    Every photon package has its own reference frame that is transformed at every scattering event in order to describe the position and direction of propagation of the photon package in the 3D model space.
    A detailed description of the mathematically and geometrically random walk of the photon packages can be found in \citet{Fischer_1993}, while an illustration of the rotation of the frame of the photon package due to scattering can be found in \cref{fig:rotation_rld}.
    POLARIS tracks the position and path of every photon package through the model space until it is absorbed, or leaves the model space and is detected by the observer.

%-----------------------------------------------------------------------

\subsection{Illuminating source}
\label{subsec:illuminating_source}

    The illuminating source is assumed to be a spherical, spatially extended radiation source with radius $R_\star$.
    To simulate the emission of radiation, four independent angles are required to specify its random starting point on the stellar surface (photosphere) and direction of emission.
    The location of emission on the photosphere is defined by the angles $\theta_1$ and $\phi_1$.
    The polar angle $\theta_1$ is defined with respect to the $z$-axis of the global model space and the azimuthal $\phi_1$ is in the $(x, y)$-plane.
    The model space is illustrated in \cref{fig:model_space}.
    In this setup, the angle $\theta_1$ has values in the range $[0, \pi]$ and $\phi_1$ in the range $[0, 2\pi]$.
    The angles $\theta_2$ and $\phi_2$ describe the direction of emission at that starting point.
    Here the polar angle $\theta_2$ is the angle between the surface normal and the direction of propagation of the photon package, and $\phi_2$ is the azimuthal angle in the surface plane.
    Thus, $\theta_2$ has values in the range $[0, \pi/2]$ and $\phi_2$ in the range $[0, 2\pi]$, respectively.
    A further detailed description can be found in \citet{Cashwell_1959} or \citet{Niccolini_2003}, among others.

    As the solid angle under which a planet is seen from the central star is very small (even in the case of hot Jupiters), we restrict the emission angles $\theta_1$, $\theta_2$, and $\phi_2$ accordingly.
    Consequently, we sample only emission locations and directions to ensure that the photon package propagates towards the planet.
    To compensate the oversampling in that range, we have to weight the net energy of the photon package.
    Weighting the energy of the photon packages is a popular method for optimizing Monte Carlo simulations \citep[see, e.g.,][]{Cashwell_1959, Yusef-Zadeh_1984, Lucy_1999, Juvela_2005, Baes_2016}.
    These restrictions are necessary to avoid inefficient computations because only a small number of photon packages actually hit the planetary atmosphere if the four angles are calculated over their full range.
    For example, for an Earth-sized planet with radius $R_\mathrm{p} = \SI{6.3781e6}{m}$ at a distance of $d_\star = \SI{1}{au}$ to its central star, the solid angle under which the planet is seen from the star is approximately \SI{5.7e-9}{sr}.
    For a hot Jupiter with a radius of $R_\mathrm{p} = \SI{7e7}{m}$ and a distance of $d_\star = \SI{0.1}{au}$, the solid angle is approximately \SI{6.9e-5}{sr}.
    
    The allowed range for the emission angles has to be determined individually for every photon package.
    While the range of the angle $\theta_1$ depends on the planetary radius $R_\mathrm{p}$, stellar radius $R_\star$, and the distance between the planet and the star $d_\star$, the range of the angle $\theta_2$ in addition depends on the position on the stellar photosphere that is defined by $\theta_1$.
    The range of the angle $\phi_2$ furthermore depends on the direction that is given by the previously sampled angle $\theta_2$.
    A description of the calculation of the allowed range for the emission angles is given in \cref{app:boundaries_emission}.

    While a constant brightness distribution of the stellar photosphere as seen from the planet is certainly a valid approximation in most cases, the impact of limb darkening might become relevant for close-in planets.
    If the stellar brightness distribution is known, it can be considered directly during the sampling of the emission angles.
    For the case of restricted emission, we can simply adjust the weight of the photon package according to the stellar intensity profile.

%-----------------------------------------------------------------------

\subsection{Planetary model}
\label{subsec:planetary_model}

\begin{figure}
        \centering
        \includegraphics[width=8.8cm]{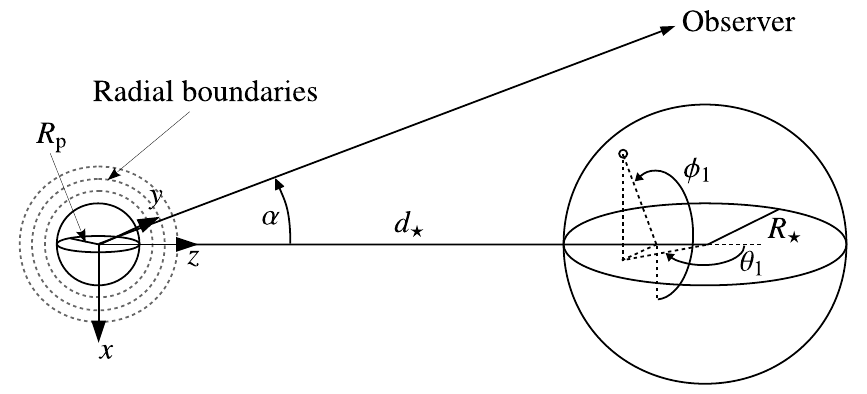}
        \caption{Illustration of the model space.
        The planet with radius $R_\mathrm{p}$ is located at the center of the model space, i.e., the origin of the coordinate system.
        The dashed lines represent the radial cell boundaries that divide the atmosphere above the surface into individual layers.
        The star, located at distance $d_\star$ along the positive $z$-axis, has a radius of $R_\star$.
        The first two emission angles are sampled and define the starting position of a photon package on the stellar surface.
        The position is described by the polar angle $\theta_1$ with respect to the $z$-axis and the azimuthal $\phi_1$ in the $(x, y)$-plane.
        The second polar angle $\theta_2$ is defined with respect to the surface normal and the azimuthal angle $\phi_2$ is defined in the surface plane at this location.}
        \label{fig:model_space}
    \end{figure}

    After the emission the photon package travels towards the planet and experiences a random optical depth $\tau$ (e.g., in the atmosphere) before it interacts.
    The scattering and absorbing properties of our model atmosphere are defined by the optical properties of the gas particles (molecules, atoms) and cloud, aerosols, or dust particles.
    As shown in \cref{fig:model_space}, the planet with radius $R_\mathrm{p}$ is in the center of the 3D model space.

%-----------------------------------------------------------------------

\subsubsection{Atmosphere}
\label{subsubsec:atmosphere}
    
    The atmospheric structure is described by a spherical grid with radial atmospheric boundaries and with polar and azimuthal boundaries.
    Therefore, both vertical and horizontal inhomogeneities of the atmosphere can be considered. 
    Inside a spherical grid cell, the number density of particles is constant.
    A simple assumption for the radial atmospheric pressure profile or density profile, and thus the optical depth of the gas phase, is based on the condition of hydrostatic equilibrium and the equation of state for the ideal gas.
    Starting at the top of the atmosphere we can describe the increase in pressure towards the center as
    \begin{equation}
    \label{eq:hydrostatic_equilibrium}
        \frac{\dd p}{\dd h} = \rho g =
        \frac{p M_\mathrm{g} g}{\mathcal{R} T} =
        \frac{p m_\mathrm{g} g}{k_\mathrm{B} T},
    \end{equation}
    where $g$ is the gravitational acceleration, $\mathcal{R}$ the gas constant, $k_\mathrm{B}$ the Boltzmann constant, and $T$ the temperature.
    The quantity $M_\mathrm{g}$ is the molar mass of the gas in \si{kg. mol^{-1}} and $m_\mathrm{g}$ is the mass of one gas particle (or molecule) in \si{kg}.
    By integrating this equation we can calculate the layer boundaries using
    \begin{equation}
        h_{i+1} = h_i - \mathcal{H}(p_{i+1}) \ln \left( \frac{p_{i+1}}{p_i} \right), \quad
        \mathcal{H} = \frac{\mathcal{R} T(p)}{M_\mathrm{g} g},
    \end{equation}
    where $\mathcal{H}$ is the scale height with a pressure-dependent temperature profile.
    Such a temperature profile could be given by the approximation of a gray atmosphere \citep[e.g.,][]{Hansen_2008, Guillot_2010}.
    The pressure bins are logarithmically distributed between $p_\mathrm{min}$ at the upper boundary (at $h_0 = \SI{0}{m}$) of our model and $p_\mathrm{max}$ at the lower boundary at $R_\mathrm{p}$.
    If we express the gas density following \cref{eq:hydrostatic_equilibrium} as $\rho = m_\mathrm{g} n$ with a number density $n$ of the gas particles, the optical depth $\tau_\mathrm{g}$ of a given layer can be expressed as
    \begin{equation}
    \label{eq:layer_optical_depth}
        \frac{\tau_\mathrm{g}}{C_\mathrm{ext}} =
        n(h_{i+1} - h_i) = \frac{p_{i+1} - p_i}{m_\mathrm{g} g},
    \end{equation}
    where $C_\mathrm{ext}$ is the extinction cross section of the gas particles (i.e., the scattering cross section plus the absorbing cross section).
    
    For the gas particles (i.e., particles that are very small compared to the wavelength) the scattering properties can be described by the Rayleigh scattering theory including depolarization due to anisotropic molecules.
    The elements of the scattering matrix for Rayleigh scattering are given by \citet{Hansen_Travis_1974}.
    The wavelength-dependent scattering cross section can be expressed by the Rayleigh cross section \citep{Sneep_2005},
    \begin{equation}
        C_\mathrm{sca}(\lambda) =
        \frac{ 24 \pi^3 }{ \lambda^4 n_\mathrm{s}^2 } \frac{ (n'(\lambda)^2 - 1)^2 }{ (n'(\lambda)^2 + 2)^2 }
        \left( \frac{ 6 + 3 \rho_\mathrm{d} }{ 6 - 7 \rho_\mathrm{d} } \right),
    \end{equation}
    where $n'(\lambda)$ is the wavelength-dependent real part of the refractive index and $n_\mathrm{s}$ is the number density at standard conditions; the density $n_\mathrm{s}$ is usually referred to as the Loschmidt constant.
    The real part of the refractive index can be approximated by
    \begin{equation}
        n'(\lambda) =
        A \left( 1 + \frac{B}{(\lambda \times \num{e6})^2} \right) + 1,
    \end{equation}
    where $A$ and $B$ are real constants and depend on the type of the considered gas \citep{Cox_2000}.
    The wavelength-, pressure-, and temperature-dependent absorption cross section $C_\mathrm{abs}(\lambda, p, T)$ can be calculated by knowing the strength and width of the absorption line of the corresponding gas \citep[see, e.g.,][]{Rothman_1998, Tennyson_2016}.
    
    In addition to the gas particles, we also consider clouds, aerosols, and dust particles in the atmosphere.
    For these larger particles the elements of the scattering matrix $\vec{F}$ as well as the particle cross sections $C_\mathrm{ext}$, $C_\mathrm{abs}$, and $C_\mathrm{sca}$ are calculated with MIEX \citep{Wolf_2004}, which is based on the Mie scattering theory and takes into consideration arbitrarily large size parameters $2\pi r / \lambda$.
    The cloud particles have a size distribution described by \citet{Hansen_1971}:
    \begin{equation}
        N(r) \propto r^{ (1-3v_\mathrm{eff}) / v_\mathrm{eff} }\ \mathrm{e}^{ -r / (r_\mathrm{eff}v_\mathrm{eff}) }.
    \end{equation}
    Here $r$ is the radius of the cloud particle, $r_\mathrm{eff}$ the effective particle radius, and $v_\mathrm{eff}$ the effective variance \citep[see][]{Hansen_1971}.
    Similar to the gas phase, the cloud particles, aerosols, or dust particles are constant in a grid cell defined by radial, azimuthal, and polar boundaries.
    
    To account for the absorption of radiation, the energy of the interacting photon package is weighted by the single scattering albedo of the particle
    \begin{equation}
        \omega = \frac{ C_\mathrm{sca} }{ C_\mathrm{ext} }
                = \frac{ C_\mathrm{sca} }{ C_\mathrm{sca} + C_\mathrm{abs} }.
    \end{equation}
    
%-----------------------------------------------------------------------
    
\subsubsection{Surface}
\label{subsubsec:surface}
    
    If a photon package reaches the planetary surface, it is reflected back into the current layer or transmitted into the next layer, depending on the properties of the surface.
    Similar to the case of scattering by a particle, the Stokes vector of the photon package is transformed using a surface reflection (or transmission) matrix to describe the change in state of polarization.
    
    For a diffuse (Lambertian) reflecting surface the reflection matrix element $R_{11}$ is equal to the planetary surface albedo, while the remaining elements are zero.
    Therefore, the reflected radiation does not depend on the incident angle and is fully depolarized.
    For a specular reflection, however, the angle of the reflected or transmitted photon package depends on the incoming angle and on the refractive indices of the optical media that are separated by the surface.
    In this case the reflection and transmission matrix is based on the Fresnel equations \citep[see, e.g.,][]{Zhai_2010, Garcia_2012}.
    
%-----------------------------------------------------------------------
\section{Selected test cases}
\label{sec:selected_test_cases}

    We start with various selected test cases to verify the correct implementation of the concepts outlined in \cref{sec:radiative_transfer_model}.
    The test cases include a homogeneous Rayleigh scattering atmosphere, a diffuse reflecting sphere, and an inhomogeneous atmosphere with both gas and cloud particles.
    Our computations are compared to the results by \citet{Buenzli_2009}, the analytical solution by \citet{Russel_1916}, and the results by \citet{Karalidi_2012} for the Rayleigh scattering atmosphere, the diffuse reflecting sphere, and the inhomogeneous atmosphere, respectively.
    To make our results independent of the general planetary and stellar parameters, we normalize the detected flux by
    \begin{equation}
    \label{eq:normalize_flux}
        \pi B(\lambda,T)\ \frac{R_\star^2\ R_\mathrm{p}^2}{d_\star^2\ d_\mathrm{obs}^2}.
    \end{equation}
    An overview of the general model parameters is given in \cref{tab:parameter_model}.
    
    \begin{table}
        \centering
        \caption{General model parameters.}
        \begin{tabular}{l c r c}
            \hline\hline
            Parameter & Symbol & Value & Model\\
            \hline
            \multirow{2}{*}{Radius: Planet} & \multirow{2}{*}{$R_\mathrm{p}$} & \SI{7.0e7}{m} & I, III\\
            & & \SI{6.3781e6}{m} & II\\
            \multirow{2}{*}{Separation:} & \multirow{3}{*}{$d_\star$} & \SI{0.1}{au} & I\\
            \multirow{2}{*}{Planet -- Star} & & \SI{1.0}{au} & II\\
            & & \SI{3.0}{au} & III\\
            Wavelength & $\lambda$ & \SI{550}{nm} & All\\
            \hline
        \end{tabular}
        \label{tab:parameter_model}
    \end{table}
    
%-----------------------------------------------------------------------
    
\subsection{Homogeneous cloud-free atmosphere}
\label{subsec:cloud-free_atmosphere}
    
    We consider a simple homogeneous atmosphere consisting of gas particles with an optical depth $\tau_\mathrm{g}$ above a diffuse reflecting surface with albedo $\omega_\mathrm{s}$, and compare our computations with the results by \citet{Buenzli_2009}.
    The grid is divided into one radial, one polar, and one azimuthal cell since the atmosphere is both vertically and horizontally homogeneous.
    As this atmospheric layer contains only gaseous particles, the scattering properties are described by the Rayleigh scattering theory.
    The depolarization factor is set to zero and we ignore the absorption by gas particles ($C_\mathrm{abs} = 0$).
    Thus, the total radial optical depth of the gaseous layer with height $H_\mathrm{atm}$ can be calculated by
    \begin{equation}
        \tau_\mathrm{g} = C_\mathrm{sca}\ n_\mathrm{g}\ H_\mathrm{atm}.
    \end{equation}
    Here, $C_\mathrm{sca}$ is the scattering cross section of the particles, and $n_\mathrm{g}$ is the number density that can be scaled to simply adjust the optical thickness of the atmosphere.
    The radius of our planetary model is set to $R_\mathrm{p} = \SI{7e7}{m}$ with a height of the atmospheric layer of $H_\mathrm{atm} = \SI{e5}{m}$.
    The distance to the star is set to $d_\star = \SI{0.1}{au}$.
    This model is referred to as model I.
    
\begin{figure}
        \centering
        \includegraphics[width=8.8cm]{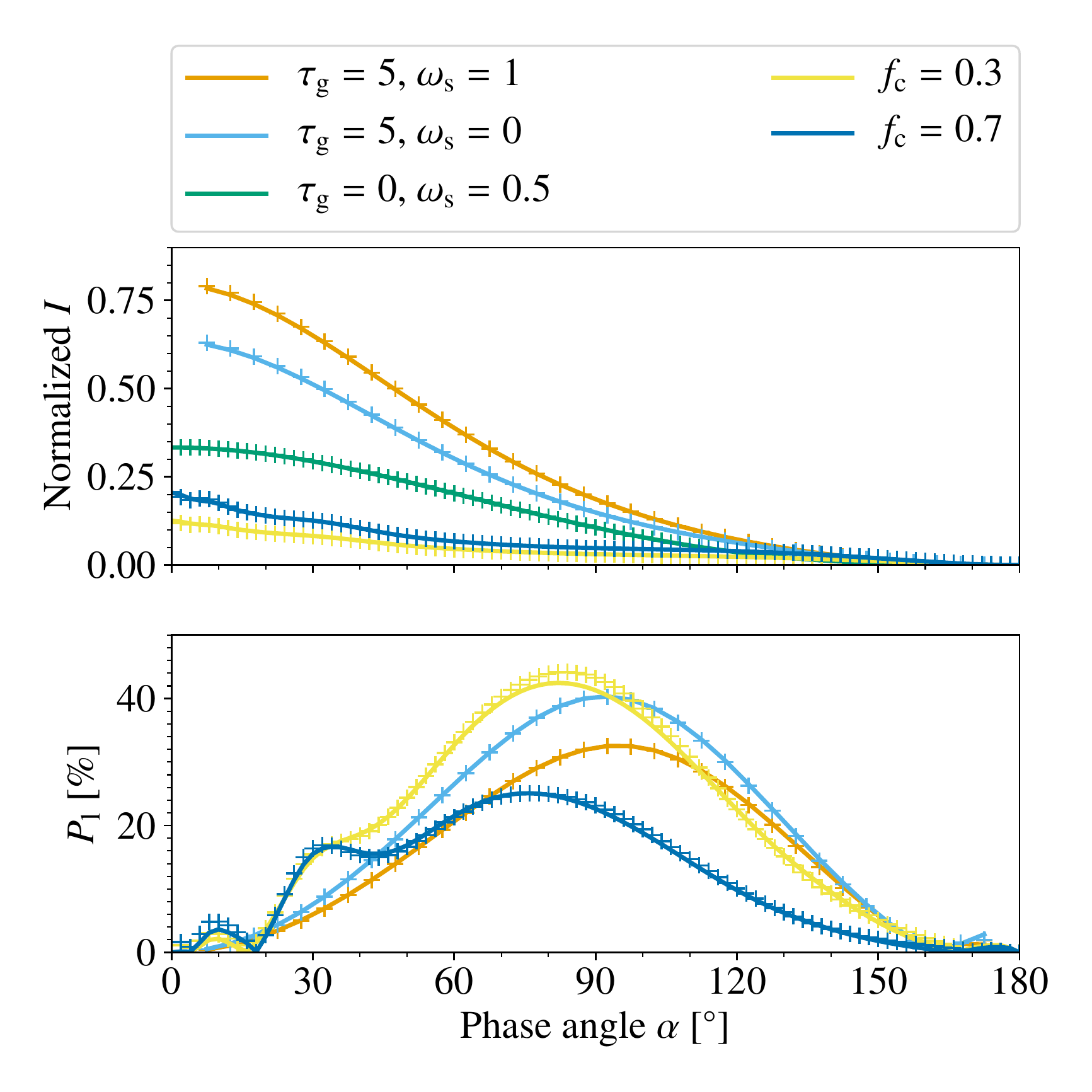}
        \caption{Normalized reflected flux (\emph{top}) and degree of linear polarization (\emph{bottom}) as a function of the phase angle $\alpha$ for various atmospheric models ($\tau_\mathrm{g} = 5, 0, 0.097$; $\omega_\mathrm{s} = 1, 0.5, 0$; $f_\mathrm{c} = 0, 0.3, 0.7$).
        Shown are the cloud-free homogeneous atmosphere (model I; orange and light blue lines),   the Lambertian reflecting sphere without an atmosphere ($\tau_\mathrm{g} = 0$;  green line), and the cloudy inhomogeneous atmosphere (model II; yellow and dark blue lines) with various cloud fractions $f_\mathrm{c}$.
        Our numerical calculations (plus symbols) are compared with results by \citet{Buenzli_2009} for the homogeneous atmosphere, with the analytical solution by \citet{Russel_1916} for the Lambertian reflecting sphere, and with the results by \cite{Karalidi_2012} for the inhomogeneous atmosphere.
        The color-coding applies to both panels.
        See \cref{subsec:cloud-free_atmosphere} and \cref{subsec:cloudy_atmosphere} for details.}
        \label{fig:test_case_1}
    \end{figure}

    \Cref{fig:test_case_1} shows the normalized reflected flux and degree of linear polarization as a function of the phase angle $\alpha$ for an optical depths $\tau_\mathrm{g} = 5$ with a surface albedo $\omega_\mathrm{s} = 0$ and $\omega_\mathrm{s} = 1$.
    We find a very good agreement for the reflected flux (relative difference $\vert \delta I \vert < \num{0.008}$ at $\alpha = \ang{7.5}$) and the degree of linear polarization (relative difference $\vert \delta P_\mathrm{l} \vert < \num{0.002}$ at $\alpha = \ang{92.5}$) between our numerical calculations and the results by \citet{Buenzli_2009}.
    
    An optical depth of $\tau_\mathrm{g} = 0$ corresponds to the case of a simple Lambertian reflecting sphere without an atmosphere above.
    Here, the computations are compared to the analytical solution derived by \citet{Russel_1916},
    \begin{equation}
    \label{eq:lambertian_reflection}
        I(\alpha) = \frac{2 \omega_\mathrm{s}}{3 \pi} ( \sin(\alpha) + (\pi - \alpha) \cos(\alpha) ),
    \end{equation}
    where $I(\alpha)$ is the normalized reflected flux of the planet at a phase angle $\alpha$.
    We find very good agreement between our numerical calculations and the analytical solution for the considered value of the surface albedo (relative difference $\vert \delta I \vert < \num{0.004}$ at $\alpha = \ang{0}$).
    
%-----------------------------------------------------------------------

\subsection{Inhomogeneous cloudy atmosphere}
\label{subsec:cloudy_atmosphere}

    In this section we investigate the impact of an inhomogeneous cloud cover in the atmosphere and compare our computations with the results by \citet{Karalidi_2012}.
    The atmosphere of the planetary model describes an Earth-like planet.
    Here the optical depth of the gas measured in radial direction is determined by given pressure and temperature profiles of a mid-latitude atmosphere tabulated by \citet{McClatchey_1972}.
    The distance to the star is set to $d_\star = \SI{1}{au}$ and the planet has a radius of $R_\mathrm{p} = \SI{6.3781e6}{m}$ with a height of the atmosphere of \SI{e5}{m}.
    The optical depth of each layer is calculated applying \cref{eq:layer_optical_depth}, the given pressure profile, and by using $M_\mathrm{g} = \SI{2.8964e-2}{kg. mol^{-1}}$, which is typical for air \citep{Cox_2000}.
    The resulting total radial optical depth of the gas particles is thus $\tau_\mathrm{g} \approx \num{0.097}$.
    We use a depolarization factor of \num{0.028} (typical for air; \citealt{Bates_1984}) and ignore the absorption by the gas particles.
    The cloud particles are located between \SI{3}{km} and \SI{4}{km} above the surface and have a total radial optical depth of $\tau_\mathrm{c} = 2$.
    They have an effective radius of $r_\mathrm{eff} = \SI{2}{\micro m}$ and an effective variance of $v_\mathrm{eff} = 0.1$ \citep{Karalidi_2012} with a refractive index of $\num{1.335} + \num{e-5} \mathrm{i}$ \citep{Karalidi_2011} typical for water droplets.
    In contrast to the gas particles, the cloud layer has a horizontally inhomogeneous structure (i.e., a random patchy pattern) that is described by the cloud fraction $f_\mathrm{c}$.
    The atmosphere is divided into 32 radial, 90 polar, and 180 azimuthal cells to define the 3D atmospheric model.
    The surface has an albedo of $\omega_\mathrm{s} = 0$, thus it absorbs all incident radiation.
    This model is referred to as model II.
    
\begin{figure}
        \centering
        \includegraphics[width=8.8cm]{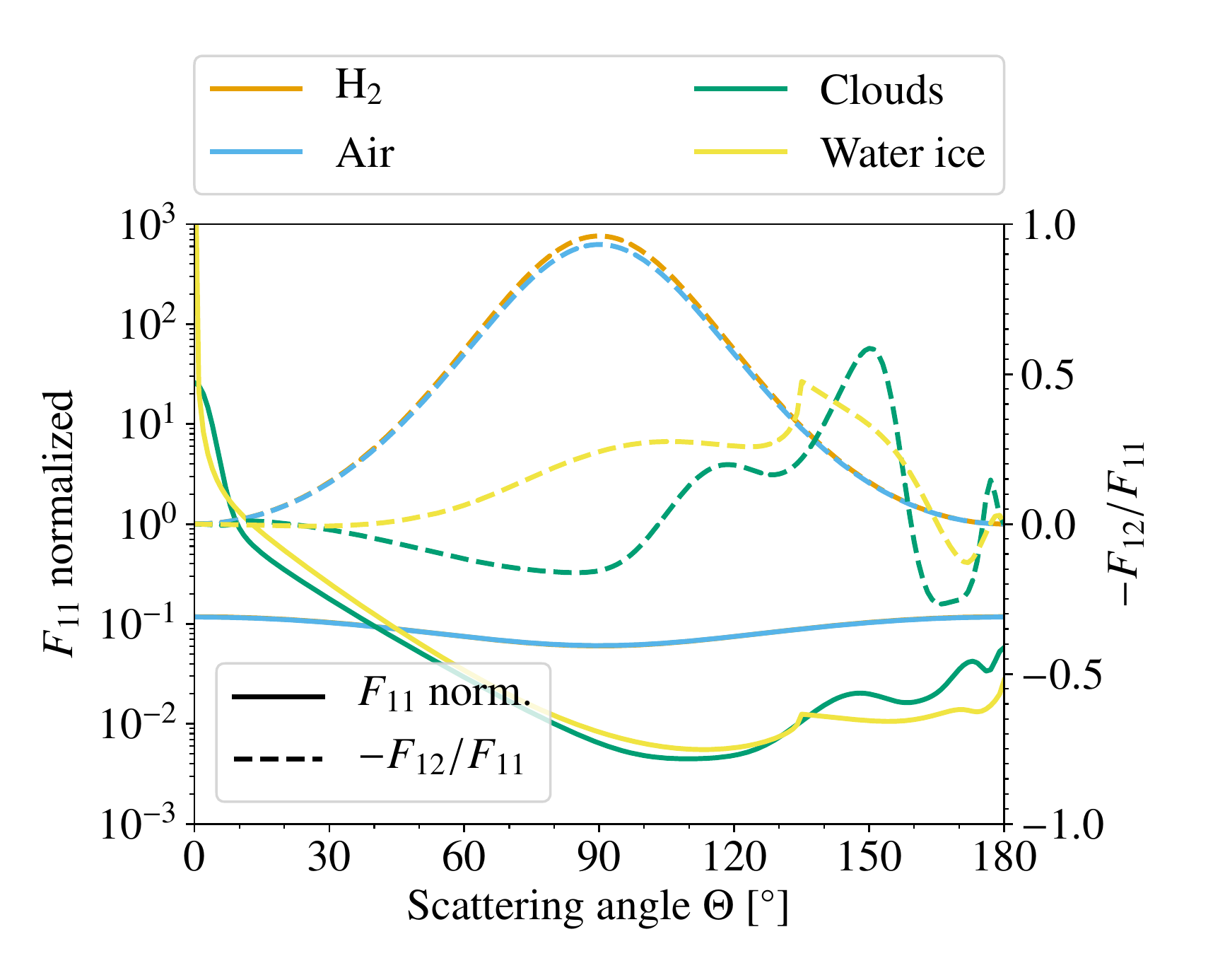}
        \caption{Normalized matrix element $F_{11}$ (solid lines) and the ratio $-F_{12} / F_{11}$ (dashed lines) as a function of the scattering angle ($\Theta = \ang{0}$: forward scattering) for molecular hydrogen (Rayleigh scattering with $\rho_\mathrm{d} = 0.02$), air (Rayleigh scattering with $\rho_\mathrm{d} = \num{0.028}$), water cloud particles (Mie scattering, see \cref{subsec:cloudy_atmosphere} for details), and water-ice particles (Mie scattering, see \cref{sec:circumplanetary_ring} for details).
        The element $F_{11}$ is normalized such that $\int F_{11}\ \dd \Omega = 1$.
        The solid orange line ($\mathrm{H}_2$) is not visible because it coincides with the solid light blue line (air).
        The plotted matrix elements of the cloud particles and of the water-ice particles are averaged over all grain sizes and are weighted by the particle size distribution using \cref{eq:averaging}.}
        \label{fig:matrix_elements}
    \end{figure}

    The derived elements $F_{11}$ and $F_{12}$ of the Müller matrix of air and for the water cloud particles as a function of the scattering angle $\Theta$ are shown in \cref{fig:matrix_elements}.
    Since we use a size distribution for the cloud particles, the plotted values are averaged over all grain sizes ranging from $r_\mathrm{min}$ to $r_\mathrm{max}$ and weighted by the size distribution $N(r)$ in order to represent the effective values:
    \begin{equation}
    \label{eq:averaging}
        F_{ij} = \frac{ \int_{r_\mathrm{min}}^{r_\mathrm{max}} F_{ij}(r) N(r)\ \dd r }{ \int_{r_\mathrm{min}}^{r_\mathrm{max}} N(r)\ \dd r }.
    \end{equation}
    For incoming unpolarized radiation, the scattering matrix element $F_{11}$ indicates  the fraction of radiation scattered towards the direction described by the phase angle $\alpha = \pi - \Theta$.
    The ratio $F_{12} / F_{11}$ indicates the linear polarization degree after single scattering for initially unpolarized light.
    In the case of $F_{12} < 0$ the polarization vectors are perpendicular to the scattering plane, while they are parallel for $F_{12} > 0$.
    While a Rayleigh scattering profile is obtained in the case of molecular hydrogen, the scattering distribution of the cloud particles strongly increases towards smaller scattering angles, indicating strong forward scattering.
    In addition, the matrix element $F_{12}$ shows a maximum at a scattering angle of approximately \ang{150} which is the characteristic rainbow feature of water droplets.

    The normalized reflected flux and the degree of linear polarization as a function of the phase angle for various cloud fractions are shown in \cref{fig:test_case_1}.
    Our numerical results are compared with the results by \citet{Karalidi_2012}.
    We find a good agreement for the reflected flux (relative difference of $\vert \delta I \vert < \num{0.01}$ at $\alpha = \ang{0}$), but some deviation for the degree of linear polarization (relative difference of $\vert \delta P_\mathrm{l} \vert < \num{0.05}$ at $\alpha = \ang{90}$) between our numerical calculations and the results by \citet{Karalidi_2012}.
    The deviation is larger for the smaller cloud fraction, so we assume that the deviation in the degree of polarization is either due to our random cloud coverage since equal cloud fractions can produce slightly different observing signals \citep{Rossi_2017} and/or due to the additional horizontal transport of radiation at the terminator if the atmosphere is optically thin.
    
%-----------------------------------------------------------------------

\section{Impact of a circumplanetary ring}
\label{sec:circumplanetary_ring}

    As a case study we investigate the impact of a circumplanetary ring on the net reflected flux and polarization.
    Circumplanetary rings are expected to be detectable through their influence on the light curve of the host star during a transit event of such a planet \citep[e.g.,][]{Barnes_2004}.
    \citet{Ohta_2009} derived specific predictions for corresponding photometric and spectroscopic signatures.
    While these studies are limited to the case of transiting planets, \citet{Sucerquia_2020} developed an analytical model to estimate the photometric signatures of non-transiting exorings due to scattering.
    However, this model is restricted to the assumption of diffuse reflection by both the planetary atmosphere and the ring, so no preferential direction of scattered light is expected.
    However, the atmosphere and the circumplanetary ring consist of particles that have characteristic scattering properties (e.g., described by Mie scattering in the case of cloud or dust particles).

    While the rings of Saturn mainly consist of water-ice \citep{Nicholson_2008}, ices only exist if the separation to the star satisfies the relation
    \begin{equation}
        d_\star \gtrsim \left( \frac{L_\star}{16\pi \sigma T_\mathrm{sub}^4} \right)^{1/2} =
        \SI{2.7}{au} \left( \frac{L_\star}{L_\odot} \right)^{1/2},
    \end{equation}
    where $L_\star$ is the luminosity of the central star, $\sigma$ the Stefan–Boltzmann constant, and $T_\mathrm{sub} = \SI{170}{K}$ the sublimation temperature of water-ice \citep{Gaudi_2003}.
    Therefore, rings of close-in planets (e.g., hot Jupiters) are expected to consist of dusty and rocky material.
    Following this line of reasoning, we choose a separation of $d_\star = \SI{3}{au}$ between the planet and the central star, and for the sake of simplicity water-ice particles as the only material in the circumplanetary ring with optical properties based on the complex refractive index taken from \citet{Warren_2008}.
    The ring ranges from \SI{7.7e7}{m} to \SI{1.5e8}{m}, measured from the center of the planet, to mimic the width of Saturn's main ring system \citep{Colwell_2009}.
    With an opening angle of \ang{;;0.2} of the ring, we satisfy the upper limit of \SI{200}{m} of Saturn's ring thickness observed by \citet{Lane_1982}.
    The particles have a power-law size distribution $N(r) \propto r^{-3.5}$, as expected to result from a collisional cascade \citep{Dohnanyi_1969}.
    The grain  radii range from \SI{0.1}{\micro m} to \SI{1}{mm}.
    In this simple model the number density of water-ice particles is constant throughout the ring.
    The derived matrix elements $F_{11}$ and $F_{12}$ of the Müller matrix of water-ice are shown in \cref{fig:matrix_elements}.
    The particles show strong forward scattering.
    The high single scattering albedo of $\omega \approx 1$ at the considered wavelength of \SI{550}{nm} leads to a low absorption by the water-ice particles.
    
    The optical depth of the ring, measured throughout the vertical coordinate of the ring at the outer edge, amounts to
    \begin{equation}
        \tau_\mathrm{r} = C_\mathrm{ext}\ n_\mathrm{r}\ H_\mathrm{r} .
    \end{equation}
    Here $n_\mathrm{r}$ is the number density of the water-ice particles and $H_\mathrm{r}$ the total height of the ring at the outer edge.
    We investigate an optically thin ring with $\tau_\mathrm{r} = 0.1$, a ring with an optical depth of $\tau_\mathrm{r} = 1$, and an optically thick ring with $\tau_\mathrm{r} = 5$.
    This corresponds to relatively low ring masses of $M_\mathrm{r} \approx \SI{2e-6}{M_\mathrm{S,r}}$, $M_\mathrm{r} \approx \SI{2e-5}{M_\mathrm{S,r}}$, and $M_\mathrm{r} \approx \SI{1e-4}{M_\mathrm{S,r}}$, respectively, with $\mathrm{M}_\mathrm{S,r} = \SI{1.54e19}{kg}$ being the mass of the rings of Saturn \citep{Iess_2019}.
    However, the low mass ratio is due to the much smaller grains compared to those in the rings of Saturn \citep{Zebker_1985}.
    
    We assume a simple single layer homogeneous atmosphere consisting of molecular hydrogen ($\mathrm{H}_2$) without any cloud particles for the planetary model.
    This model is referred to as model III.
    The scattering properties of molecular hydrogen are described by the Rayleigh scattering theory with a depolarization factor of 0.02 \citep{Hansen_Travis_1974}.
    We only consider  scattering at gas particles and ignore absorption.
    The planetary radius is set to $R_\mathrm{p} = \SI{7e7}{m}$ with a height of the atmosphere of $H_\mathrm{atm} = \SI{e5}{m}$.
    The optical depth of the atmosphere is set to $\tau_\mathrm{g} = 5$ above a diffuse reflecting surface with an albedo of $\omega_\mathrm{s} = 1$.
    
    Taking various inclinations of the ring $i$ with respect to the orbital plane of the planet into account, we first consider an observer who is, as before, located in the plane of the planetary orbit (edge-on orbit).
    Subsequently, we choose an observer whose line of sight is perpendicular to the planetary orbit (face-on orbit).
    As half of the planetary disk is illuminated throughout the entire orbit of the planet, any variations in the (polarized) reflected flux are due to the circumplanetary ring.
    In addition, the polarization degree resulting from scattering in the atmosphere of the planet is largest since the scattering angle (in the case of single scattering) is approximately \ang{90} towards the observer.
    
%-----------------------------------------------------------------------
\subsection{Edge-on orbit}
\label{subsec:ring_edge_on}

\begin{figure}
        \centering
        \includegraphics[width=8.8cm]{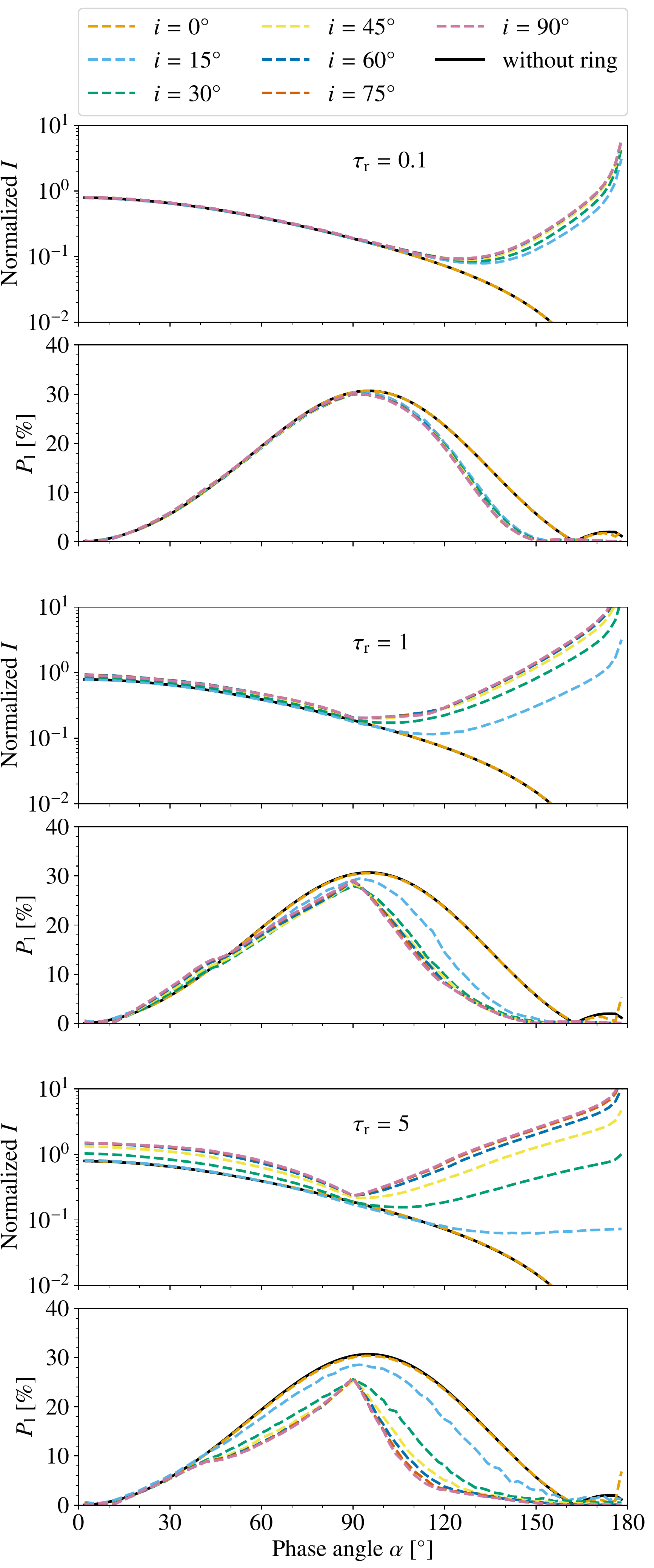}
        \caption{Normalized reflected flux $I$ and degree of linear polarization $P_\mathrm{l}$ of the planet--ring system for various inclination angles $i$ (see \cref{subsec:ring_edge_on} for details).
        The planetary orbit is seen edge-on by the observer.
        Following cases are presented: an optically thin ring with $\tau_\mathrm{r} = 0.1$ (\emph{top} pair), a ring with an optical depth $\tau_\mathrm{r} = 1$ (\emph{middle} pair) and an optically thick ring with $\tau_\mathrm{r} = 5$ (\emph{bottom} pair).
        The solid black line pertains to a planet without a ring.
        The color coding applies to all figures.}
        \label{fig:circumplanetary_ring_edge_on}
    \end{figure}

    \Cref{fig:circumplanetary_ring_edge_on} shows the normalized reflected flux $I$ and the degree of linear polarization $P_\mathrm{l}$ of the planet--ring system as a function of the phase angle $\alpha$.
    As reference, the reflected flux and degree of linear polarization of the planet without a ring is represented as a solid black line.
    
    For an inclination $i = \ang{0}$ the ring is in the plane of the orbit (i.e., seen edge-on by the observer).
    As the ring has a small vertical extension, the reflected flux and degree of linear polarization is similar to that of the planet alone.
    This is independent of the optical depth as well.
    However, if the ring is inclined, the illuminating and viewing conditions change.
    For all inclination angles, the strongest impact of the ring on the reflected flux is found at larger phase angles.
    Here, the reflected flux $I$ strongly increases.
    While the reflected radiation from the planet is very small at large phase angles, the ring has a strong impact due to the forward scattering of water-ice particles.
    In addition, the degree of linear polarization decreases since the degree of polarization is small after single scattering at water-ice particles (see \cref{fig:matrix_elements}).
    If the ring faces the observer ($i = \ang{90}$), the deviation of the reflected flux and degree of linear polarization compared to the case of a planet without a ring is largest because the observed surface area is largest at this inclination angle.
    
    At small phase angles, the reflected radiation is dominated by radiation scattered in the planetary atmosphere.
    Thus, the reflected flux and degree of linear polarization is compareable to the case without a circumplanetary ring.
    While for an optically thin ring ($\tau_\mathrm{r} = 0.1$) most of the incoming radiation penetrates the ring, the amount of back-scattered radiation increases with increasing optical depth ($\tau_\mathrm{r} = 1$ and $\tau_\mathrm{r} = 5$).
    Thus, the reflected flux increases at smaller phase angles.
    In addition, the degree of linear polarization $P_\mathrm{l}$ decreases due to the increased likelihood of multiple scattering combined with a small polarization after single scattering at water-ice particles.
    
\begin{figure*}
        \centering
        \includegraphics[width=18cm]{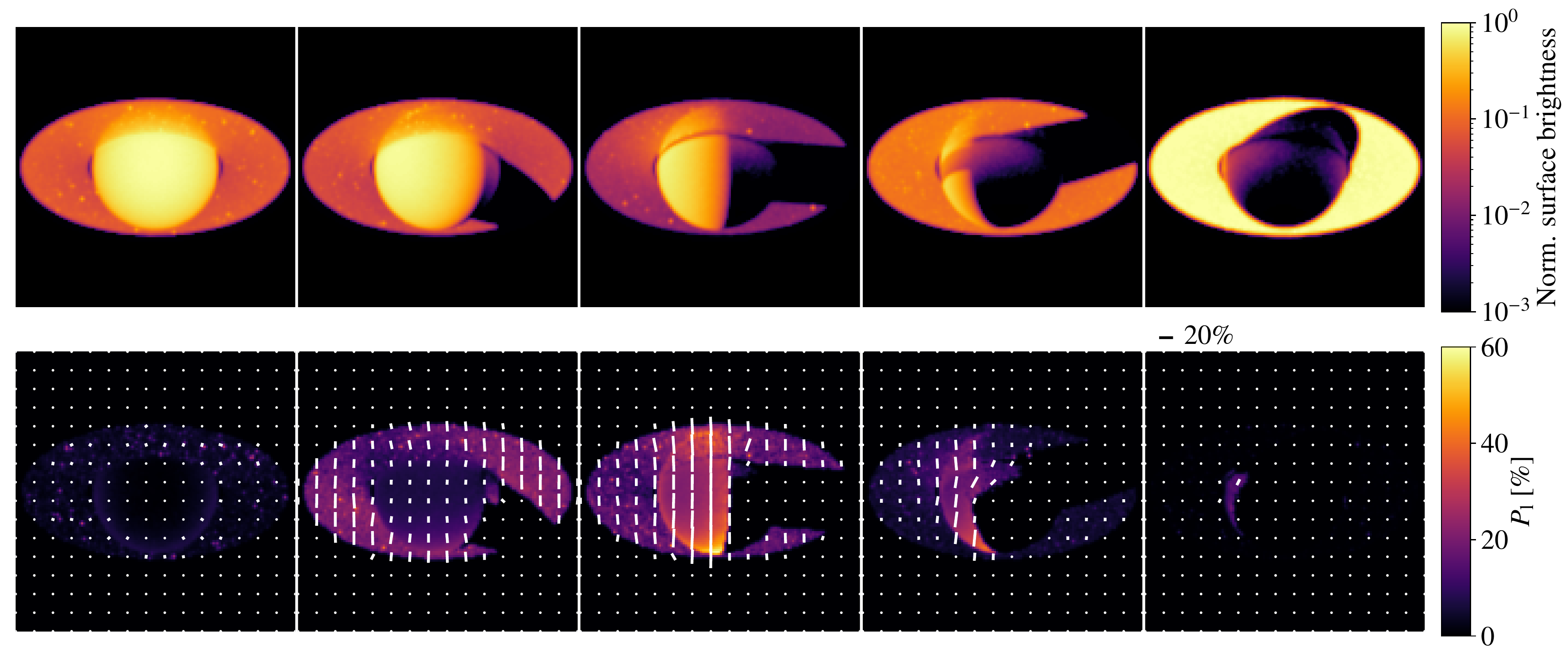}
        \caption{Spatially resolved maps of the reflected flux (\emph{top}) and degree of linear polarization (\emph{bottom}) for phase angles \ang{0}, \ang{40}, \ang{80}, \ang{120}, and \ang{160} (from \emph{left} to \emph{right}).
        The direction and length of the white vectors in the lower figures represent the angle and degree of the linear polarization, respectively.
        The vectors are oriented perpendicular to the incoming radiation since the matrix element $F_{12}$ is negative for both molecular hydrogen and water-ice particles at these scattering angles (see \cref{fig:matrix_elements}).
        The scale of the vector length is displayed on the right side between the figures.
        The ring has an inclination of $i = \ang{30}$ and an optical depth of $\tau_\mathrm{r} = 1$.
        The planetary orbit is seen edge-on by the observer.
        See \cref{subsec:ring_edge_on} for details.}
        \label{fig:resolved_circumplanetary_ring_edge_on}
    \end{figure*}

    In addition to the properties of the particles, shadowing effects have an impact on the reflected (polarized) flux as well (see \cref{fig:resolved_circumplanetary_ring_edge_on} where spatially resolved images for $i = \ang{30}$ and $\tau_\mathrm{r} = 1$ are shown).
    As the planet orbits the central star, the illuminated areas of planet and ring change, and thus the net polarization of the reflected flux changes as well.
    The planet always casts shadows on the circumplanetary ring if the inclination angle is small enough.
    In addition, the ring casts shadows on the planetary body (shown for $\alpha = \ang{120}$ in  \cref{fig:resolved_circumplanetary_ring_edge_on}), resulting in a lower reflected flux.
    Furthermore, the shadow of the ring also covers parts of the planetary body, and thus lowers the degree of linear polarization, for example at $\alpha = \ang{90}$ where the polarization is at its maximum due to the Rayleigh scattering in the atmosphere.

%-----------------------------------------------------------------------
\subsection{Face-on orbit}
\label{subsec:ring_face_on}
    
\begin{figure}
        \centering
        \includegraphics[width=8.8cm]{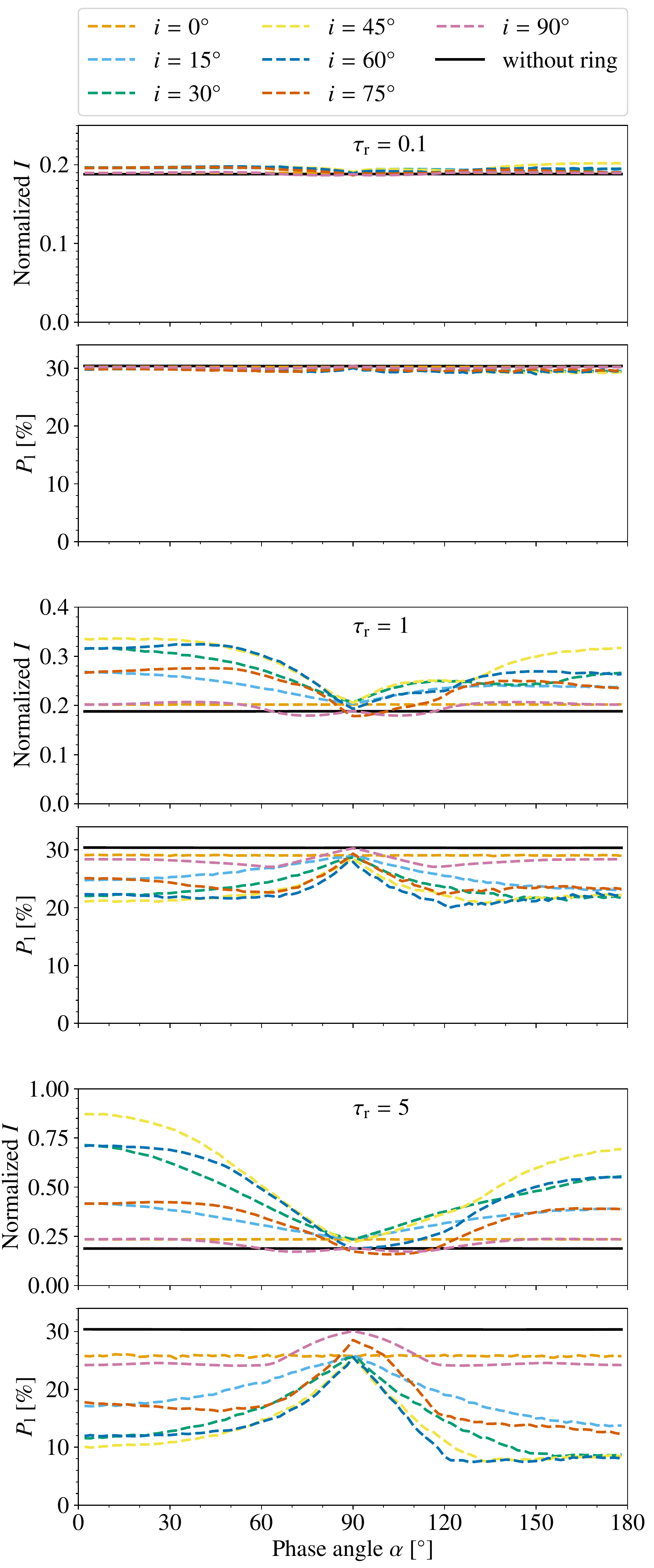}
        \caption{Similar to \cref{fig:circumplanetary_ring_edge_on}, but the planetary orbit is seen face-on by the observer (see \cref{subsec:ring_face_on} for details).
        The color coding applies to all figures.}
        \label{fig:circumplanetary_ring_face_on}
    \end{figure}

    In \cref{fig:circumplanetary_ring_face_on} the normalized reflected flux $I$ and the degree of linear polarization $P_\mathrm{l}$ of the planet--ring system are shown.
    Similar to the previous case, we consider different inclination angles $i$ of the ring with respect to the orbital plane of the planet and for different optical depths $\tau_\mathrm{r}$.
    As reference the reflected flux and degree of linear polarization of the planet without a ring is represented as a solid black line.

    For an optically thin ring ($\tau_\mathrm{r} = 0.1$ in our model setup), the difference of the reflected flux $I$ and degree of linear polarization $P_\mathrm{l}$ compared to a planet without a ring becomes negligible.
    The radiation reflected by the planet dominates the flux since most radiation penetrates the ring without scattering.
    Subsequently, the impact on the net flux is negligibly small for this geometrical setup combined with an optically thin ring.
    
    For the case of $\tau_\mathrm{r} = 1$ and $\tau_\mathrm{r} = 5$, the reflected flux increases at smaller and larger phase angles.
    In contrast, the degree of polarization decreases.
    At $\alpha = \ang{90}$, the reflected flux and degree of linear polarization have its minimum and maximum, respectively, because the ring is illuminated from the side.
    Thus, the impact on the reflected (polarized) flux is lowest.
    This is also true if the ring is not inclined ($i = \ang{0}$).
    Here, both $I$ and $P_\mathrm{l}$ are constant throughout the entire orbit of the planet.
    In addition, the reflected flux is higher at small phase angles compared to larger phase angles.
    This is because in our geometrical setup the radiation has to penetrate the ring at larger phase angles, which lowers the observed flux with increasing optical depth.
    
\begin{figure*}
        \centering
        \includegraphics[width=18cm]{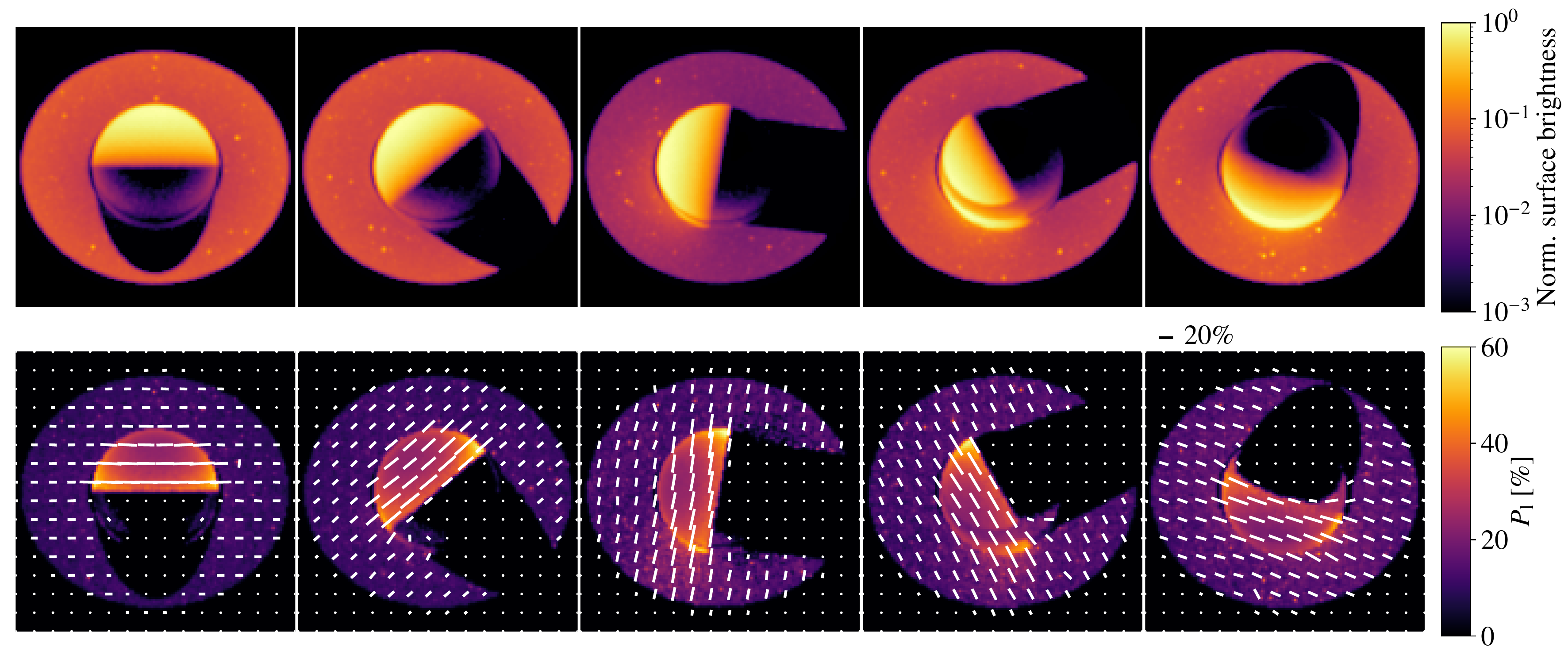}
        \caption{Similar to \cref{fig:resolved_circumplanetary_ring_edge_on}, but the planetary orbit is seen face-on by the observer.
        See \cref{subsec:ring_face_on} for details.}
        \label{fig:resolved_circumplanetary_ring_face_on}
    \end{figure*}

    \Cref{fig:resolved_circumplanetary_ring_face_on} shows spatially resolved images for $i = \ang{30}$ and $\tau_\mathrm{r} = 1$.
    As mentioned before, the reflected polarized radiation of the planet itself does not change while it orbits the central star.
    However, similar to the previous case, the planet can cast shadows on the circumplanetary ring, and vice versa.
    In addition, the total reflected flux and degree of polarization also decrease if the ring covers the planetary body.
    
%-----------------------------------------------------------------------
\subsection{Discussion}
\label{subsec:ring_discussion}
    
    The polarization contrast $C_\mathrm{pol}$, which is the ratio of the polarized flux to the total stellar flux \citep[see, e.g.,][]{Hunziker_2020}, depends on the planetary radius $R_\mathrm{p}$ and the distance to the central star $d_\star$: 
    \begin{equation}
        C_\mathrm{pol} = P(\alpha) I(\alpha) \frac{R_\mathrm{p}^2}{d_\star^2}.
    \end{equation}
    Here $I(\alpha)$ is the reflectivity (i.e., the normalized reflected intensity) and $P(\alpha)$ the degree of polarization.
    For our considered model (i.e., $R_\mathrm{p} = \SI{7e7}{m}$ and $d_\star = \SI{3}{au}$) the ratio $R_\mathrm{p} / d_\star$ is approximately \num{2e-8}.
    Including polarization, the polarization contrast amounts to \num{e-9}, which is out of reach for existing polarimetric devices \citep[e.g., SPHERE;][]{Buezit_2019}.
    So far, only close-in planets (e.g., hot Jupiters) would produce a sufficiently high polarization contrast.
    In this case the observed polarized flux can reveal circumplanetary rings of extra-solar planets.
    However, rings around these close-in planets are expected to have relatively short lifetimes because multiple forces are at work \citep[see, e.g.,][]{Goldreich_1982}, such as the radiation pressure (Poynting-Robertson effect).
    
    In addition to geometrical parameters characterizing the planet--ring and/or planet--star system investigated here, the reflected polarized flux of the circumplanetary ring also depends on the properties of the dust phase constituting the ring; however, this is outside the scope of the present study.
    
%-----------------------------------------------------------------------

\section{Summary}
\label{sec:summary}
    
    In this study we presented a numerical tool that simulates the 3D polarized radiative transfer in planetary atmospheres based on pressure and temperature profiles; various light scattering species such as various types of gas, condensates, and dust; as well as the planetary surface that is absorbing or reflecting, and (de)polarizing the incoming radiation.
    The simulation software is based on the publicly available radiative transfer code POLARIS \citep{Reissl_2016} and has been equipped with dedicated numerical routines described in this paper.
    Our approach avoids the various simplifications commonly used in previous studies, for example the locally plane-parallel planetary atmosphere, the missing cross-talk between latitudinal and longitudinal regions, a point-like star, or plane-parallel illumination.
    In particular, we are able to consider a spatially extended radiation source and take an inhomogeneous brightness distribution into account, for example due to the effect of limb darkening or stellar spots.
    In addition to describing vertical and horizontal inhomogeneities in the structure of the atmosphere and on the surface, scattering material in the local planetary environment, such as planetary rings and exomoons can also be included.
    Furthermore, the code has been optimized to handle very small planetary cross sections.
    
    We tested our numerical concepts and calculated the reflected and polarized flux for different types of planetary models.
    This includes an atmosphere-free planet with a Lambertian reflecting surface only, planets with a purely gaseous atmosphere considering various optical depths and surface albedos, and a planet with additional cloud particles in the atmosphere.
    The numerical results show very good agreement with already existing (analytically derived) computations for these atmospheric models.
    
    Furthermore, we presented a study in which we investigated the influence of a circumplanetary ring.
    We considered the cases of an observer located in the plane of the planetary orbit and an observer located perpendicular to it.
    This study shows that an additional circumplanetary ring can have an essential impact on the observed polarized radiation.

    For the transiting planet, the flux strongly increases at larger phase angles due to the forward scattering of water-ice particles.
    However, the degree of linear polarization decreases at these phase angles because the polarization degree after single scattering at water-ice particles is small.
    This is also true if the ring is not inclined ($i = \ang{0}$).
    Here, the reflected flux and degree of linear polarization is similar to the case of a planet without a ring.
    
    If the orbit of the planet is seen face-on by the observer, then a change in the reflected (polarized) flux is due only to the ring.
    The reflected (polarized) flux for a ring optical depth of $\tau_\mathrm{r} = 0.1$ differs  slightly from the results of the same planetary model without a ring.
    For an inclined ring with an optical depth of $\tau_\mathrm{r} = 1$ and $\tau_\mathrm{r} = 5$, however, the reflected flux increases while the degree of linear polarization decreases.
    The characteristic profile is due to the additional scattering inside the ring and various shadowing effects on the planet and the ring as well.
    However, if the ring is not inclined, the observed polarized radiation remains constant as the planet orbits the star.
    
%-----------------------------------------------------------------------

\begin{acknowledgements}
    R.B. thanks the DFG for financial support under contract WO857/18-1.
    We thank the anonymous referee for very useful suggestions.
\end{acknowledgements}

\bibliographystyle{aa}
\bibliography{bibliography.bib}

%-----------------------------------------------------------------------

\begin{appendix}

\section{Boundaries for restricted emission}
\label{app:boundaries_emission}
    
    The planet is located in the origin of the model space, while the radiation source is located on the positive $z$-axis.
    Thus, the first polar angle $\theta_1$ is restricted from
    \begin{equation}
        \theta_\mathrm{1,min} =
            \pi - \arccos \left( \frac{R_\star - R_\mathrm{p}}{d_\star} \right)
    \end{equation}
    to $\pi$.
    For the second polar angle $\theta_2$, we determine the angle $\theta_\mathrm{c}$ that is needed to rotate to the center of the planet, i.e, the center of the model space (see \cref{app:angles_of_rotation} for the corresponding calculation).
    The allowed range of values is then given by
    \begin{equation}
        \theta_\mathrm{2,min} = \theta_\mathrm{c} - \delta\theta, \quad
        \theta_\mathrm{2,max} = \theta_\mathrm{c} + \delta\theta,
    \end{equation}
    where $\delta \theta$ is the deviation depending on $R_\mathrm{p}$ and $d_\star$ with $\sin(\delta \theta) = R_\mathrm{p} / d_\star$.
    
\begin{figure}
        \centering
        \includegraphics[width=8.8cm]{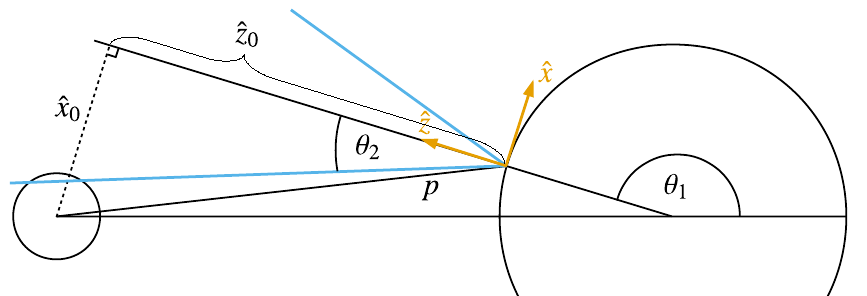}
        \caption{Photon package position (defined by $\theta_1$ and $\phi_1$) on the stellar surface and distance $p$ to the center of the planet.
        The emission cone (light blue line) with opening angle $\theta_2$ intersects with the planet in the range $[\phi_\mathrm{2,min}, \phi_\mathrm{2,max}]$.
        The new coordinate space at the location of the photon package is indicated in orange.
        See \cref{app:boundaries_emission} for details.}
        \label{fig:restricted_emission_theta_2}
    \end{figure}

\begin{figure}
        \centering
        \includegraphics[width=8.8cm]{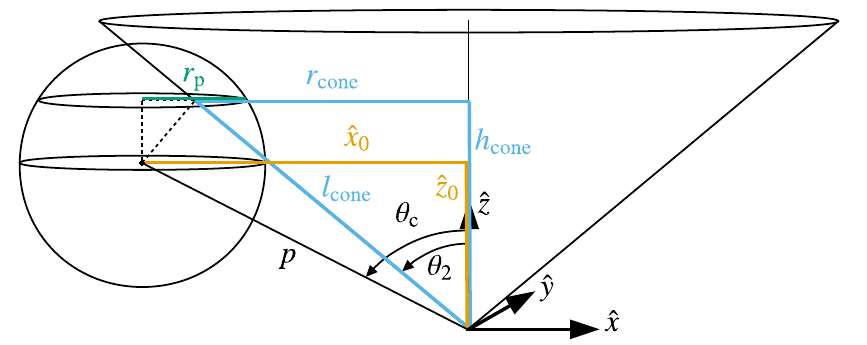}
        \caption{Intersection of the emission cone with opening angle $\theta_2$ and the planet's sphere.
        The maximum range of the azimuthal angle $\phi_2$ is at the height $h_\mathrm{cone}$ of the new coordinate space where the planet has a radius $r_\mathrm{p}$ and the cone a radius  $r_\mathrm{cone}$.
        See \cref{app:boundaries_emission} for details.}
        \label{fig:restricted_emission_cone}
    \end{figure}

    For the second azimuthal angle $\phi_2$, we set a new coordinate system $(\hat{x}, \hat{y}, \hat{z})$ originating at the location of the emission point, i.e., the apex of the emission cone, with the $\hat{z}$-axis parallel to the surface normal of the stellar surface and the $\hat{x}$- and $\hat{y}$-axis perpendicular to the surface normal.
    In \cref{fig:restricted_emission_theta_2} this coordinate space is indicated in orange.
    The $\hat{y}$-axis points into the drawing plane.
    In this new coordinate system the center of the planet is at $(\hat{x}_0, \hat{y}_0, \hat{z}_0) = (p \sin \theta_\mathrm{c}, 0, p \cos \theta_\mathrm{c})$, where $p = \vert \vec{p} \vert$ and $\vec{p}$ is the position vector of the origin of the new coordinate system in the global 3D model space.
    The height $h_\mathrm{cone}$ of the cone where the azimuthal angle $\phi_2$ is largest, the radius $r_\mathrm{cone}$ at this height, and the length of the surface line $l_\mathrm{cone}$ at this height (see \cref{fig:restricted_emission_cone}) can be calculated by
    \begin{align}
        l_\mathrm{cone} &= p \cos(\theta_\mathrm{c} - \theta_2),\\
        r_\mathrm{cone} &= l_\mathrm{cone} \sin(\theta_2),\\
        h_\mathrm{cone} &= l_\mathrm{cone} \cos(\theta_2).
    \end{align}
    The radius of the planet at this height is given by
    \begin{equation}
        r_\mathrm{p}^2 = R_\mathrm{p}^2 - (h_\mathrm{cone} - \hat{z}_0)^2.
    \end{equation}
    The 3D problem is now reduced to a 2D problem where we have to consider an intersection of a circle with radius $r_\mathrm{cone}$ with a circle of radius $r_\mathrm{p}$ in the $(\hat{x}, \hat{y})$-plane of the new coordinate space:
    \begin{align}
    \label{eq:circle_cone} 
        \hat{x}^2 + \hat{y}^2 &= r_\mathrm{cone}^2,\\
    \label{eq:circle_planet}
        (\hat{x} - \hat{x}_0)^2 + \hat{y}^2 &=
        \hat{x}^2 - 2 \hat{x} \hat{x}_0 + \hat{x}_0^2 + \hat{y}^2 = r_\mathrm{p}^2.
    \end{align}
    Putting \cref{eq:circle_cone} into \cref{eq:circle_planet} leads to
    \begin{equation}
        r_\mathrm{cone}^2 - r_\mathrm{p}^2 + \hat{x}_0^2 = 2 \hat{x} \hat{x}_0.
    \end{equation}
    By using polar coordinates in the new coordinate space with $\hat{x} = r_\mathrm{cone} \cos(\varphi)$, the equation can be rewritten as
    \begin{equation}
        \cos(\varphi) =
            \frac{r_\mathrm{cone}^2 - r_\mathrm{p}^2 + \hat{x}_0^2}{2 r_\mathrm{cone} \hat{x}_0},
    \end{equation}
    where $\pm \varphi$ is the maximum deviation of the azimuthal angle $\phi_2$.
    Finally, the boundaries for the azimuthal angle are
    \begin{equation}
        \phi_\mathrm{2,min} = \phi_\mathrm{c} - \varphi, \quad
        \phi_\mathrm{2,max} = \phi_\mathrm{c} + \varphi.
    \end{equation}
    Similar to the polar case, $\phi_\mathrm{c}$ is the angle to rotate to the center of coordinate space (see \cref{app:angles_of_rotation}).

%-----------------------------------------------------------------------

\section{Determining the angles of rotation}
\label{app:angles_of_rotation}
    
    If the direction of the incoming photon package and the direction of the outgoing photon package are given, we can calculate the angles that are needed to perform this rotation.
    This calculation is necessary, for example, to rotate to the center of coordinate space ($\phi_\mathrm{c}$ and $\theta_\mathrm{c}$, see \cref{app:boundaries_emission}).
    
\begin{figure}
        \centering
        \includegraphics[width=8.8cm]{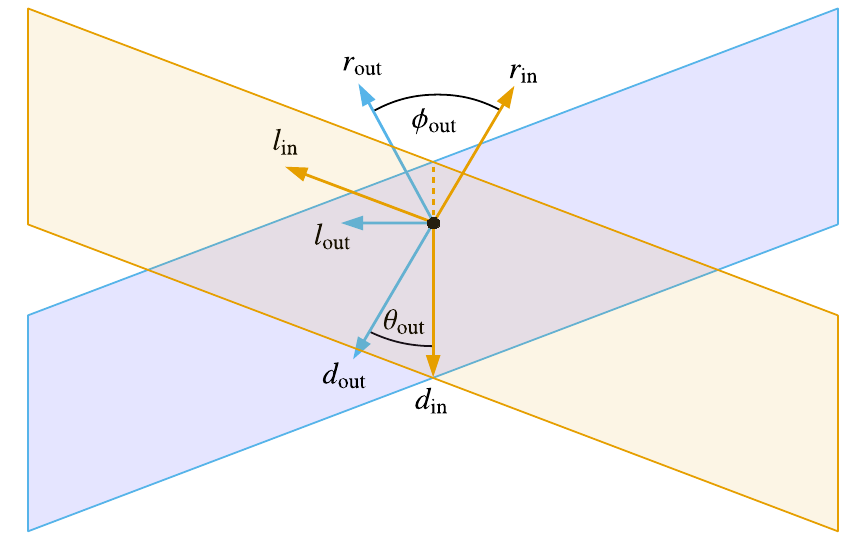}
        \caption{Rotation of the frame of the photon package before the scattering event (orange) into the frame after the scattering event (light blue).
        The planes represent the planes that are perpendicular to the $\vec{r}$-axis (i.e., the scattering planes).
        See \cref{app:angles_of_rotation} for details.}
        \label{fig:rotation_rld}
    \end{figure}

    \Cref{fig:rotation_rld} shows the rotation of the incoming $(\vec{r}_\mathrm{in}, \vec{l}_\mathrm{in}, \vec{d}_\mathrm{in})$ photon package frame into the outgoing $(\vec{r}_\mathrm{out}, \vec{l}_\mathrm{out}, \vec{d}_\mathrm{out})$ photon package frame.
    The angle $\theta_\mathrm{out}$ can be calculated by using
    \begin{equation}
        \cos\theta_\mathrm{out} = \vec{d}_\mathrm{in} \cdot \vec{d}_\mathrm{out}.
    \end{equation}
    The azimuthal angle $\phi_\mathrm{out}$ is determined by two equations due to the full range from $0$ to $2\pi$.
    With the relation
    \begin{equation}
        \vec{r}_\mathrm{out} = -\frac{\vec{d}_\mathrm{in} \times \vec{d}_\mathrm{out}}{\vert \vec{d}_\mathrm{in} \times \vec{d}_\mathrm{out} \vert},
    \end{equation}
    the first equation is
    \begin{equation}
        \sin\phi_\mathrm{out}
            = ( \vec{r}_\mathrm{in} \times \vec{r}_\mathrm{out} ) \cdot \vec{d}_\mathrm{in}
            = -\frac{\vec{r}_\mathrm{in} \cdot \vec{d}_\mathrm{out}}{\vert \vec{d}_\mathrm{in} \times \vec{d}_\mathrm{out} \vert}.
    \end{equation}
    The second equation is
    \begin{equation}
        \cos\phi_\mathrm{out}
            = \vec{r}_\mathrm{in} \cdot \vec{r}_\mathrm{out}
            = \frac{\vec{l}_\mathrm{in} \cdot \vec{d}_\mathrm{out}}{\vert \vec{d}_\mathrm{in} \times \vec{d}_\mathrm{out} \vert}.
    \end{equation}
    
\end{appendix}

\end{document}